\documentclass[aps,onecolumn,showpacs,superscriptaddress]{revtex4}
\usepackage{amsmath}
\usepackage{epsf}
\usepackage{epsfig}

\begin{document}

\title{Infrared Spectrum and STM images of Cyclohexene-2-Ethanamine: First Principle Investigation}
\author{Tekin \.Izgi}
\affiliation{Anadolu
University, Departments of Physics, Eskisehir, Turkey}
\author{Ethem Akt\"urk}\thanks{Corresponding Author}
\email[E-mail: ]{eakturk@hacettepe.edu.tr} \affiliation{Hacettepe
University, Departments of Physics Engineering, Beytepe, 06800
Ankara, Turkey}
\author{O\u{g}uz G\"ulseren}
\affiliation{Bilkent
University,Departments of Physics , 06800 Ankara, Turkey}
\author{Mustafa \c{S}enyel}
\affiliation{Anadolu
University, Departments of Physics, Eskisehir, Turkey}
\begin{abstract}

We have investigated the structure of cyclohexene-2-ethanamine molecule both theoretically and 
experimantally. Theoretical investigation is based on a first principle technique Density 
Functional Theory (DFT) using plane wave basis sets and ultrasoft pseudo-potentials while 
the experimental technique is infrared (IR) spectroscopy. Exchange-correlation potential of DFT 
was approximated in the frame of both local density approximation (LDA) and generalized gradient 
approximation (GGA) schemes.  
Vibrational
properties of this molecule are given by the assignments in the range for wavenumber $4000-400$ $cm^{-1}$. 
Stable equilibrium structure of the molecule was also obtained by using LDA and GGA. 
Obtained optimized geometrical structure was used to calculate vibrational properties and STM 
images. A remarkable
agreement was obtained between theory and experiment, especially in the
symmetric and asymmetric vibrations of NH groups.  

\noindent {\it Keywords:} Cyclohexene-2-ethanamine, First Principle, IR spectrum, STM images.

\end{abstract}

\pacs{31.15.Ar, 31.15.Ew, 33.20.Ea, 71.15.Mb } \maketitle

\section{Introduction}
\label{intro}

The design and synthesis of strong organic bases have long been an
active field of research~\cite{Hibbert,Staab,Alder1,Alder2}.
Infrared spectroscopy is  a valuable tool in order to obtain information about the molecular 
structure and properties of the molecules. This technique is
used widely  in qualitative and quantitative molecular
analysis. IR spectrum of interatomic vibrations can be used as
structural probes for determining weak changes of structure or
chemical bonding in molecules. Cyclohexene-2-ethanamine molecule
consists of cyclohexene $C_6H_{10}$ group attached to the carbon of
 ethylamine $(C_2H_7N)$. There are previous works on the cyclohexene and ethylamine structures.
 Some studies showed that the lowest energy conformations of cyclohexene are in a
 half-chair form and a boat structure.
 Basically, the cyclohexene ring can interconvert from one twisted form to the other
 over the boat conformation with $C_s$ symmetry~\cite{Rodin}.
 The point symmetry group for trans-ethylamine ion is $C_s$ whereas there is no such symmetry
for gauge-ethylamine~\cite{Zeroka}.
Cyclohexene-2-ethanamine (CyHEA) has also important industrial applications, that is used as chemical
 intermediate in rubber industry. They demonstrated prototypical non-conjugated olefinic substrate
 CyHEA which was not only a highly active substrate but also a mechanism-based inhibitor for
DBM. CyHEA was also used as a substrate and oxidizing agent for Ru
complex. Sirimanne and May reported that dopamine $\beta$-monooxygenase
(DBM) catalyzed stereo-selective allylic hydroxylation of
CyHEA~\cite{Sirimanne}. CyHEA was first synthesized by \.Izgi et
al.~\cite{Izgi} and some of IR and NMR properties of this compound were reported by them. 

Density functional theory(DFT) is a widely used and very precise \textit{ab initio} technique which is used 
to provide vibrational frequencies of organic
compounds perfectly~\cite{Handy1,Handy2,Stephens,Devlin,Lee1,Lee2}. The vibrational modes and STM images
of this molecule have not been investigated by an \textit{ab initio} theoretical method. In this
study, the molecule has been investigated by using planewave pseudopotential
 calculation based on DFT. Exchange-correlation potential of DFT scheme was taken into account within the 
LDA and GGA which are the commonly used approximations and both are used in calculation process. 
The stable conformation of the molecule is obtained by following a relaxation 
procedure within the framework of DFT under periodic boundary conditions.

\begin{figure}[htbp]
\centering \epsfig{file=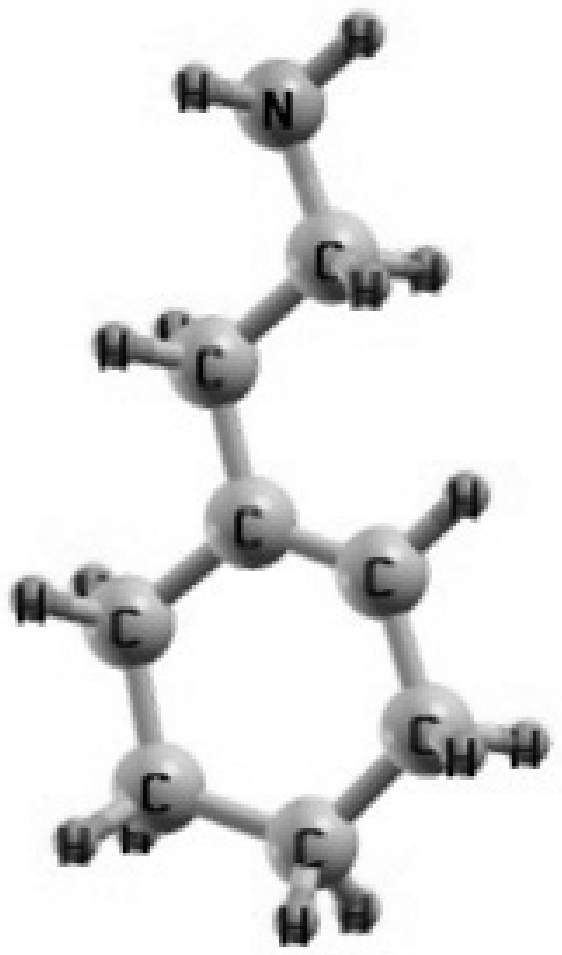, width=5cm,height=6cm}
\caption{The stable configuration of the
Cyclohexene-2-ethanamine.}\label{eps1}
\end{figure}

\section{Method}
The normal modes and STM images of the molecule were calculated with both LDA and GGA by using the freely available 
DFT program PW-SCF (Plane Wave Self Consistent Field) ~\cite{pwscf} which uses plane wave basis sets for electronic 
wavefunctions. For all calculations, we have used
Perdew-Zunger  ~\cite{PZ} and Perdew-Burke-Ernzerhof ~\cite{PBE} exchange-correlation parameterizations for LDA and GGA, respectively 
and Vanderbilt~\cite{Vanderbilt} ultrasoft pseudopotentials. The electronic wavefunctions
were expanded in terms of plane waves with kinetic energy cut-off
up to 25 Ry. The special k-points of the molecule in the cubic cell is selected as $q=0$ gamma point. The lattice constant of cubic cell is 20 bohr(au). 

For experimental work, the pure Cyclohexene-2-ethanamine in liquid form was obtained from Aldrich Chemical Co.,
USA and was used without further purification.
The IR spectra of the molecule in liquid form was recorded to be in the range of $4000-400$ $cm^{-1}$ using 
Perkin Elmer FT-IR 2000 spectrometer with a resolution of $4$ $cm^{-1}$.

\section{Results and Discussions}
\begin{figure}[htbp]
\centering \epsfig{file=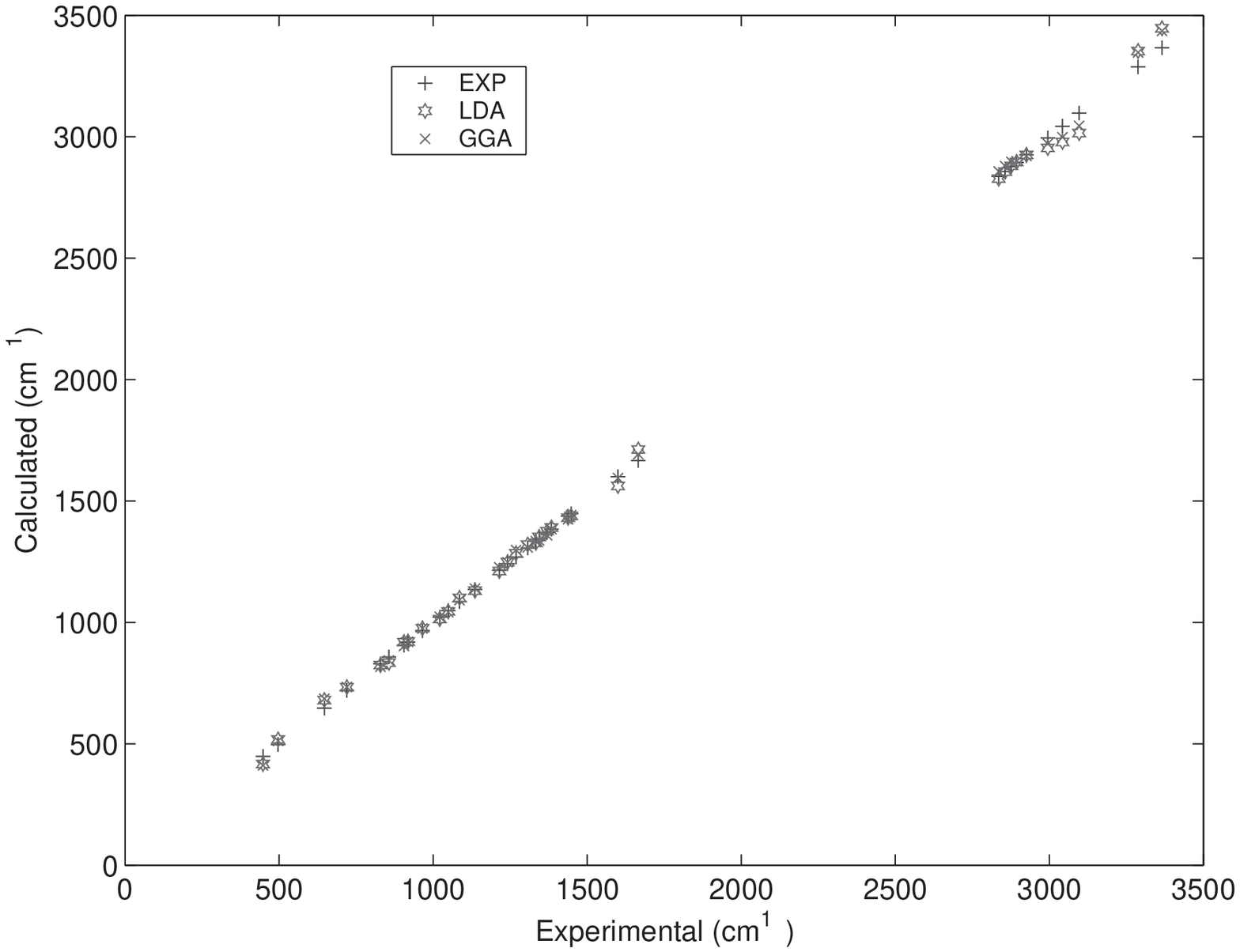, width=8cm,height=6cm}
\caption{The comparison of experimental data with GGA and LDA
results.}\label{eps2}
\end{figure}

The calculated stable structure of CyHEA is shown
in Fig.\ref{eps1} which was drawn by XCrySDen (Crystalline Structures and Densities) program~\cite{xcrysden}.
The vibrational assignments and frequencies of
cyclohexene-2-ethanamine was reported experimantally by \.Izgi et al.~\cite{Izgi}.
The spectral properties of the molecule were evaluated through the
calculated vibrational frequencies of the free ligand molecule.
The calculated and experimental infrared spectra data of the
molecule are given in Table.\ref{table1}. The experimental, GGA and
LDA results are also compared in Fig.\ref{eps2}.

\begin{table*}
\caption{Experimental and theoretical values of vibration
frequencies of cyclohexene-2-ethanamine. }\footnote{ Ass.,
assignments; Exp., experimental; Freq., frequency; v, very; s,
strong; m, medium; w, weak, sh, shoulder; b, broad; str,
stretching; bend, bending; sciss, scissoring; twist, twisting;
wag, wagging; s, symmetric; a, asymmetric.}\label{table1}
\begin{tabular}{c c c c c c}
\hline\hline
mode&\multicolumn{2}{c}{Experimental~\cite{Izgi}}&\multicolumn{2}{c}{Calculated}\\
\hline
 &Assignments &$IR  freq.$&LDA& GGA&
\\ $\nu_1$&$N-H a-str $& 3366s&    3447 & $3436 $&
\\ $\nu_2$&$N-H s-str $& 3288s&    3353 & $3347$&
\\ $\nu_3$&$\nu_{23}    $& 3097vw&   3015 & $3045$&
\\ $\nu_4$&$\nu_{1}     $& 3043m &   2979 &$2998$&
\\ $\nu_5$&$C-H str (CH3)$&2995m  &   2954 &$2971 $&
\\ $\nu_6$&$\nu_{2}+C-H str (CH_3)$& 2926vs &  2925 &$2921 $&
\\ $\nu_7$&$\nu_{25}+C-H str (CH_2)$&2894vw &   2896 &$ 2899$&
\\ $\nu_8$&$\nu_{26}$&2877vw &   2880 &$2897$&
\\ $\nu_9$&$\nu_{27}+C-H str (CH_3)$&2857vs &   2856 &$ 2879$&
\\ $\nu_10$&$\nu_{5}+C-H str (CH_3)$&2836vs &   2827 &$2857$&
\\ $\nu_{11}$&$\nu_{6}$& 1666 m &   1712 &$1693 $&
\\ $\nu_{12}$&$NH_2 sciss$& 1600mb &   1562 &$ 1594$&
\\ $\nu_{13}$&$CH_2 sciss$& 1505vw &   - &$ -$&
\\ $\nu_{14}$&$C-H bend (CH_3)$& 1473vw &   - &$1464 $&
\\ $\nu_{15}$&$\nu_{28}+C-H bend (CH_3)$& 1448vw &    1439&$1442 $&
\\ $\nu_{16}$&$\nu_{8}$& 1438s &    1431 &$ 1425$&
\\ $\nu_{17}$&$CH_2 wag$& 1384w &   1391 &$1384 $&
\\ $\nu_{18}$&$\nu_{9}+C-H bend (CH_3)$& 1370vw &   1376 &$ 1359$&
\\ $\nu_{19}$&$\nu_{10}$& 1344m &   1352 &$ 1338$&
\\ $\nu_{20}$&$\nu_{30}$& 1334w &   1328 &$1334 $&
\\ $\nu_{21}$&$NH_2 twist$& 1307w &   1320 &$ 1309$&
\\ $\nu_{22}$&$\nu_{32}$& 1269m &   1285 &$1298$&
\\ $\nu_{23}$&$\nu_{11}+CH_2 twist$& 1242w &   1248 &$1247$&
\\ $\nu_{24}$&$\nu_{12}$& 1215w &   1212 &$1227$&
\\ $\nu_{25}$&$\nu_{34}$& 1136m &   1131 &$1139$&
\\ $\nu_{26}$&$(C-C,C-N) a-str$& 1086w &   1101&$1092$&
\\ $\nu_{27}$&$\nu_{15}$& 1066w &    -&$1080$&
\\ $\nu_{28}$&$\nu_{35}$& 1049w &   1045&$1042$&
\\ $\nu_{29}$&$\nu_{36}+CH_3 rock$& 1022w &   1014 &$1023$&
\\ $\nu_{30}$&$\nu_{16}$& 966w &   976 &$975$&
\\ $\nu_{31}$&$\nu_{37}$& 919m &    921 &$920$&
\\ $\nu_{32}$&$\nu_{17}$& 906vw &   919 &$903 $&
\\ $\nu_{33}$&$\nu_{38}$& 857w &   834 &$843$&
\\ $\nu_{34}$&$\nu_{18}+CH_2 rock$& 829m &   824 &$818$&
\\ $\nu_{35}$&$\nu_{19}NH_2 wag$& 800m &   - &$804$&
\\ $\nu_{36}$&$\nu_{39}$& 720sh &   733 &$734$&
\\ $\nu_{37}$&$\nu_{40}$& 647w &   680 &$686$&
\\ $\nu_{38}$&$\nu_{20}$& 497vw &   518 &$512$&
\\ $\nu_{39}$&$\nu_{41}$& 448w &   417 &$411$&
\\
\hline
\end{tabular}
\end{table*}

\begin{figure}[htbp]
\centering \epsfig{file=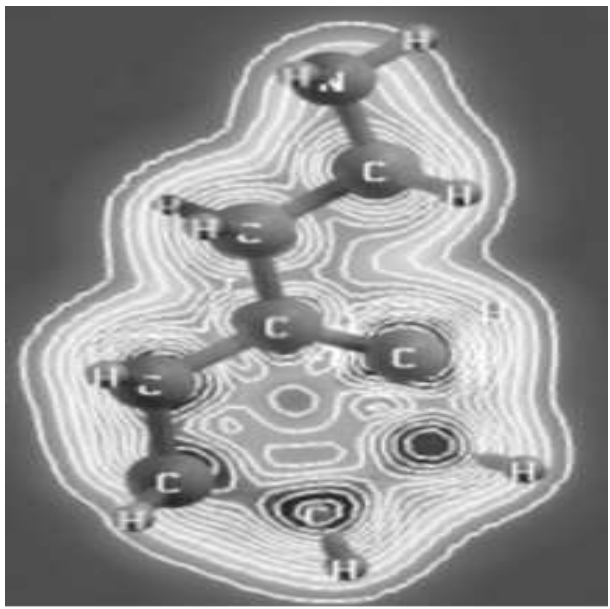, width=8cm,height=8cm}
\caption{Calculated STM images for HOMO.}\label{eps4}
\end{figure}

\begin{figure}[htbp]
\centering \epsfig{file=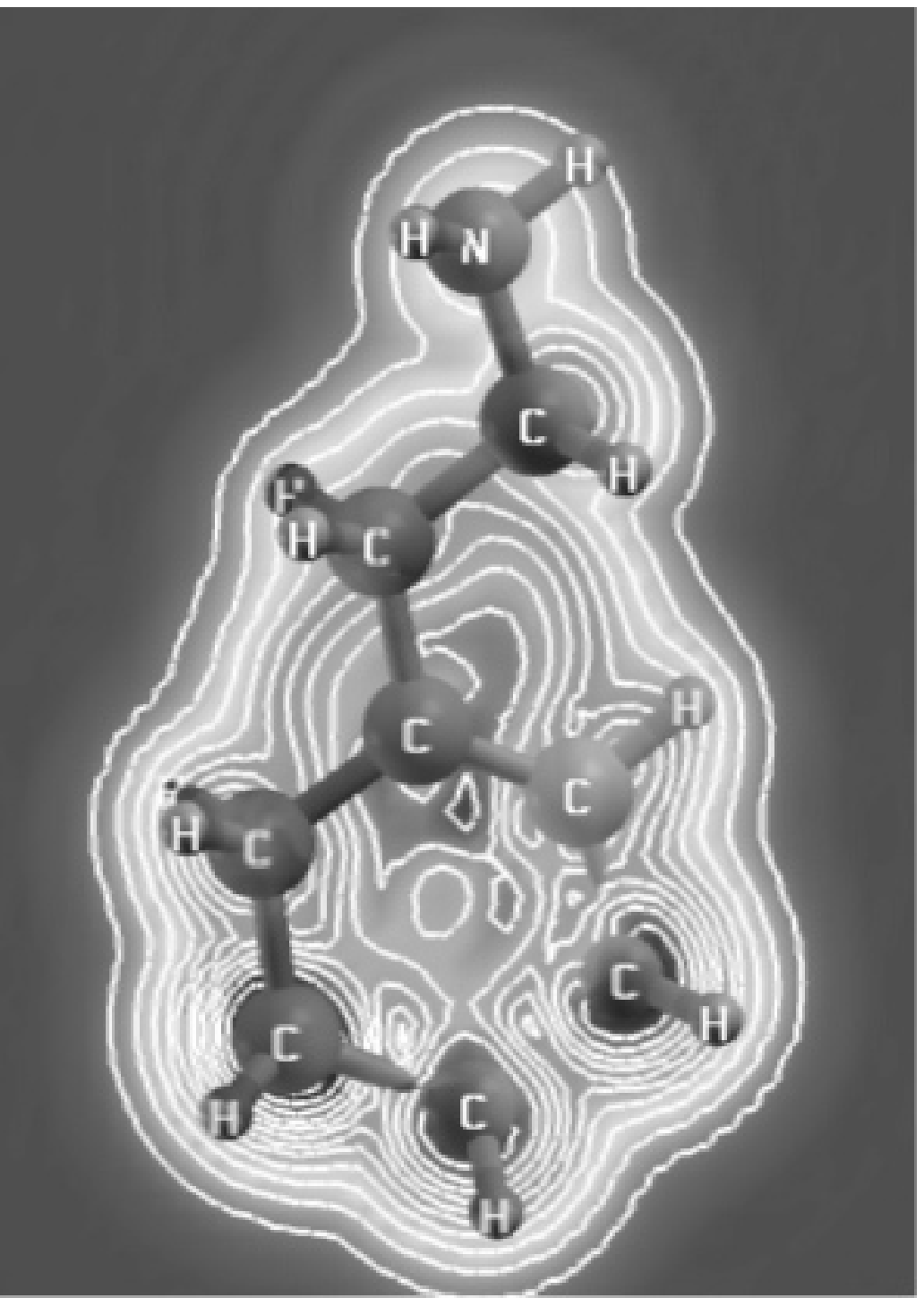, width=8cm,height=8cm}
\caption{Calculated STM images for LUMO.}\label{eps5}
\end{figure}

 The strong N-H asymmetric and symmetric stretch bands seen in Table.\ref{table1} are due to the contribution of ethylamine (see Fig.\ref{eps4}). C-H stretch bands  between $3000-3100$ $cm^{-1}$ are
attributed to cyclohexene group and the very strong C-H stretch
bands at $2926$ $cm^{-1}$ and $2835$ $cm^{-1}$ result from ethylamine. The very strong bands are attributed to the attachment of ethylamine and cyclohexene and appear between $2830-2920$ $cm^{-1}$.
Most of the modes below the $1300$ $cm^{-1}$ arise from cyclohexen. If
the vibrational assignments of the molecule involving these groups
are investigated, it is seen that the assignments obtained for the
molecule  also involve the group frequencies. Furthermore,
the observed medium broad band appears at $829$ $cm^{-1}$ is an N-H bending
band as well as a group frequency. There is also a good
agreement between the experimental and the theoretical vibrational
frequencies in the region of $4000-400$ $cm^{-1}$ except some
GGA and LDA results.
 The ground state energy of the molecule was obtained to be -128.66 ryd and -128.53 ryd for GGA and
 LDA, respectively.
 Finally, we examined the electronic
 properties by using calculated STM images for cyclohexene-2-ethanamine. In
 Fig.\ref{eps4} and Fig.\ref{eps5} which were drawn by using XCrySDen, we calculated the STM images at
 constant current and bias voltage -2.5 eV and 2.5 eV, respectively.
 These results supply a microscopic model for STM images and can serve as a
 source for STM experiments for organic molecules.

\begin{figure}[htbp]
\centering \epsfig{file=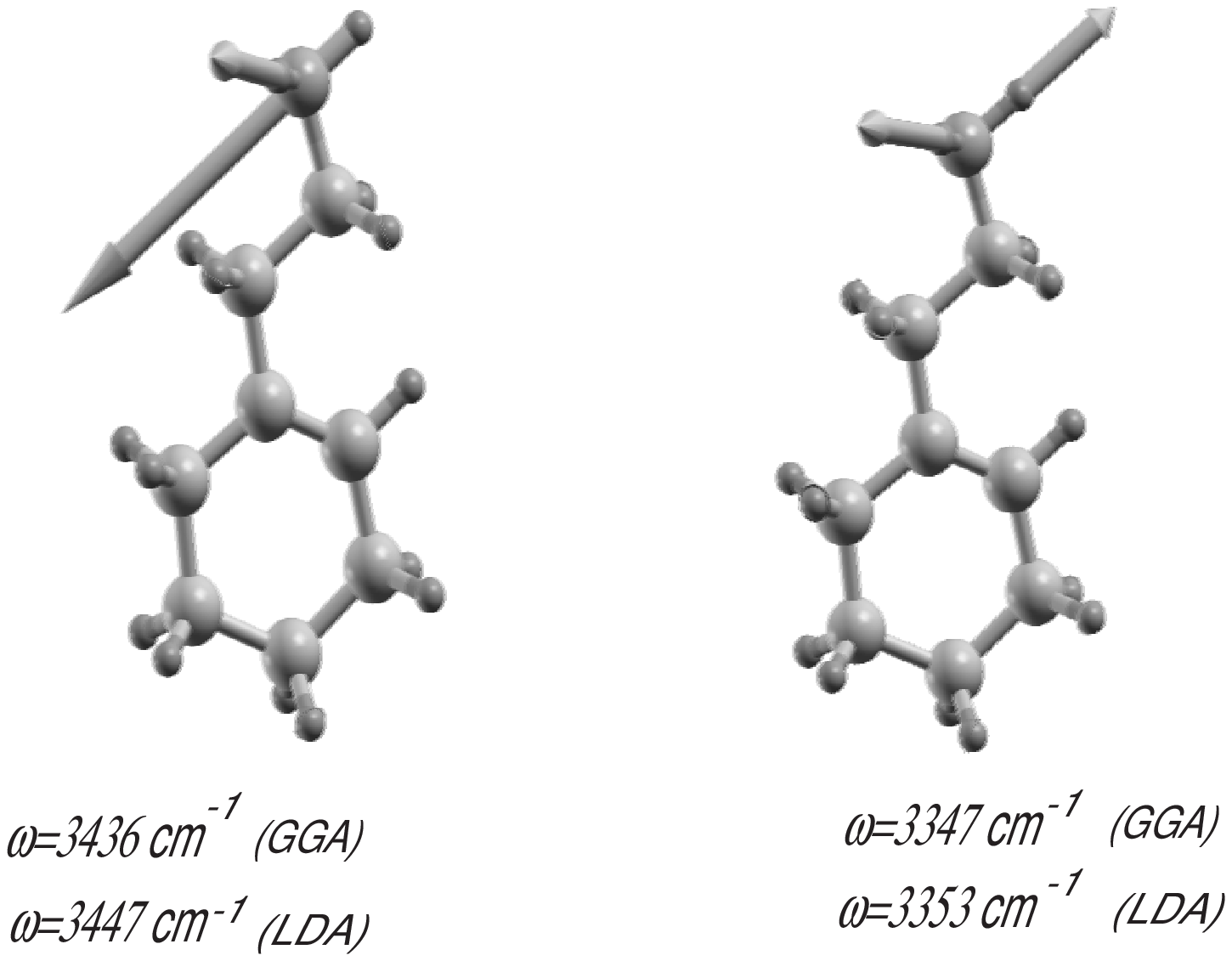, width=8cm,height=8cm}
\caption{Calculated  N-H asymmetric and symmetric
stretch.}\label{eps3}
\end{figure}

\section{Conclusion}
The experimental and the theoretical investigation of CyHEA
molecule have been performed successfully by using FT-IR and
density functional theory calculations. For all calculations, it
is shown that the results of GGA and LDA methods are in excellent
agreement with all experimental findings. Thus, density functional
theory (DFT) methods are suitable for the calculation of ground
state properties and potential energies. Hence, DFT is
an excellent method for calculating vibrational spectra and STM images from first
principles.
\section{ACKNOWLEDGEMENTS}
We thank G.Gokoglu and T. Boz for improving our paper English.

\label{secupdate}

\end{document}